# Semi-Supervised Segmentation of Functional Tissue Units at the Cellular Level

Volodymyr Sydorskyi [1], Igor Krashenyi [2], Denis Sakva [3] and Oleksandr Zarichkovyi [1]

[1]   *National Technical University of Ukraine "Igor Sikorsky Kyiv Polytechnic Institute", 37, Prosp. Peremohy, Kyiv, 03056, Ukraine*
[2]   *Ukrainian Catholic University, Ilariona Svjentsits'koho St, 17, Lviv, 79000, Ukraine*
[3]   *Dragon Capital*

**Abstract**
We present a new method for functional tissue unit segmentation at the cellular level, which utilizes the latest deep learning semantic segmentation approaches together with domain adaptation and semi-supervised learning techniques. This approach allows for minimizing the domain gap, class imbalance, and captures settings influence between HPA and HubMAP datasets. The presented approach achieves comparable with state-of-the-art-result in functional tissue unit segmentation at the cellular level. The source code is available at https://github.com/VSydorskyy/hubmap_2022_htt_solution

**Keywords** [1]
semantic segmentation, functional tissue unit, semi-supervised learning

## 1. Introduction

It is estimated that the human body contains approximately 37 trillion cells, and comprehending the complex relationships and functions among them poses a significant challenge for researchers, requiring a colossal effort [1]. One of the research directions aims to map human body at a cellular level to detect functional tissue units (FTU). FTU is defined as a unit consisting of a three-dimensional block of cells centered around a capillary, such that each cell in this block is within diffusion distance from any other cell in the same block [2]. These cellular compositions - cell population neighborhoods are responsible for performing an organ's main physiologic functions. Functional tissue units, such as colonic crypts, renal glomeruli, alveoli, etc. (examples can be observed in Figure 1) have pathobiological relevance that are essential for modeling and comprehending the development of a disease. However, manually annotating FTUs is time consuming and costly. At the same time current algorithms suffer from poor generalizability and low accuracy [3]. So the task for the competition was to segment FTU on stained microscope slides in a way that is invariant to different staining protocols. In this paper a new method is proposed, which utilizes the latest deep learning semantic segmentation [4] approaches together with domain adaptation techniques and semi-supervised learning techniques.

## 2. Related work

One of the most common approaches to functional tissue units segmentation, specifically kidney glomerulus and colon crypt [3] segmentation is based on the use of supervised learning techniques and were introduced in the previous Kaggle competition [5]. In these methods, the training data consists of annotated images, where each pixel is labeled as belonging to a particular cell or background. These techniques typically require a large amount of labeled data to achieve high accuracy, which can be time-consuming and expensive to obtain.

Most of these models are heavily inspired by the U-Net [6], UnetPlusPlus [7], FPN architectures [8], and DeepLabV3+ [9] in a combination with ImageNet pre-trained backbones such as resnet50_32x4d, resnet101_32x4d and RegNet [10]. Models used a combination of general data augmentation techniques such as flipping, rotation, scale shifting, artificial blurring, CutMix [11] and MixUp [12] to improve model performance. Models were trained using binary cross-entropy and Lovász Hinge loss [13] functions, RAdam [14], Lookahead [15], AdamW [16], SGD [17], and Adam [18] optimizers. These models used a dynamic sampling approach to sample tiles of size 512x512, 768x768 and 1024x1024 pixels from regions with visible glomeruli based on the annotations.

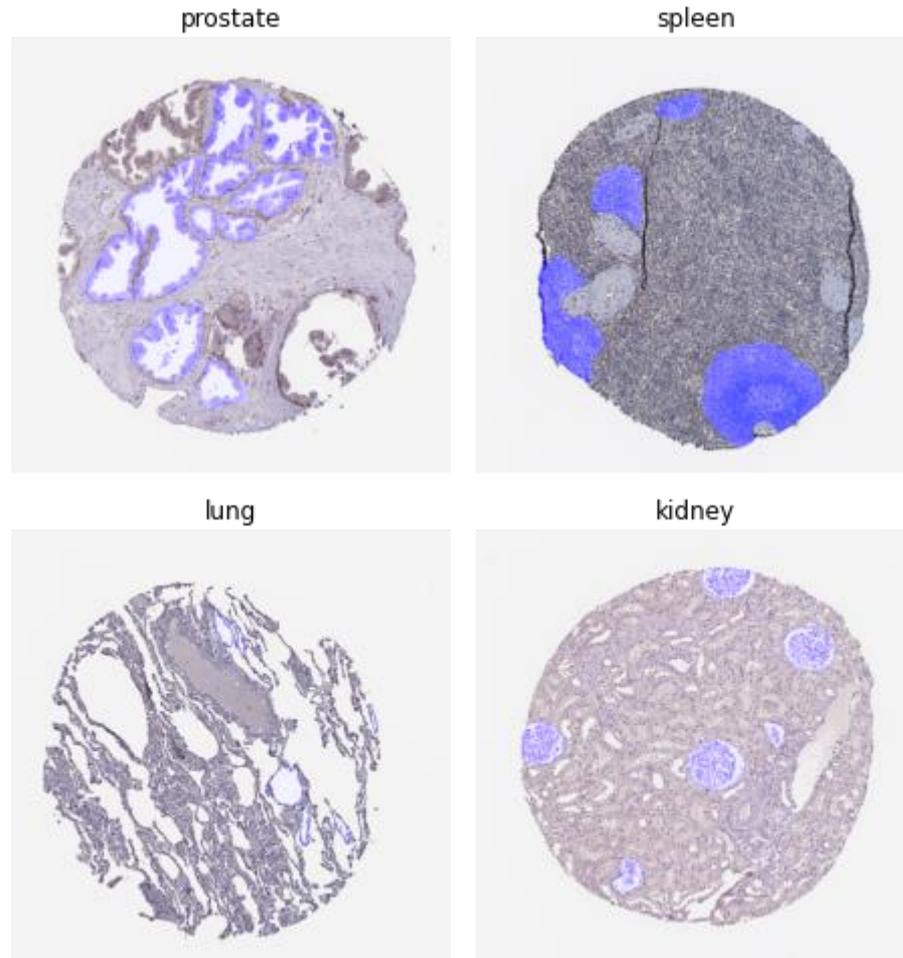

**Figure 1**: Examples of microscopic images of different organs with FTU regions highlighted with blue color

## 3. Dataset

The dataset includes biopsy slides from several organs, namely kidney, prostate, large intestine, spleen, and lung. The key feature of the proposed dataset is that it consists of images from two data sources: HPA [19-26] and HuBMAP [27]. Furthermore, the training data includes only HPA samples, while the test data comprises a mixture of HPA and HuBMAP samples [1]. Additionally, only HubMAP data was used for the final score (private dataset). The images from the HPA and HuBMAP data sources differ in staining protocol, pixel sizes, and sample thicknesses [1]. Figure 2 provides an example that illustrates the visual differences between HPA and HubMAP images. The whole slide images in the HPA and HuBMAP data sources were stained using three distinct protocols. HPA samples were stained with antibodies visualized with 3,3'-diaminobenzidine (DAB [28]), counterstained with hematoxylin, whereas HuBMAP samples were stained using either Periodic acid-Schiff (PAS [29]), hematoxylin and eosin stains (H&E [30]). Each of the staining protocols highlights different cellular structures using colored dyes, and the final stained slide images vary greatly in color, contrast, and overall image structure, making direct matching of cellular structures between images less straightforward (see Figure 4). Another crucial feature of the proposed dataset is that HubMAP images have different pixel sizes for different organs, while for HPA, it is constant (see Table 1) [1].

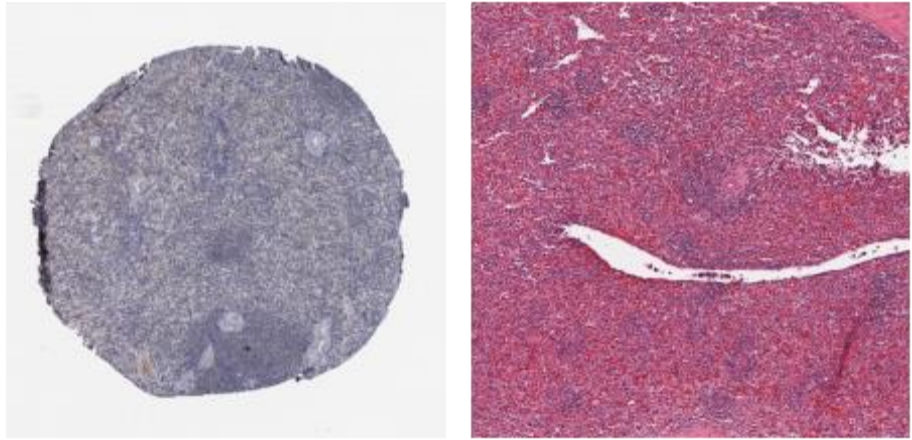

**Figure 2**: Left image refers to the spleen sample from the HPA dataset and the right to the spleen image from the HubMap dataset

**Table 1**
Pixel size table

| Organ | Dataset | Pixel size |
| --- | --- | --- |
| Kidney | HPA | 0.4 |
| Large intestine | HPA | 0.4 |
| Lung | HPA | 0.4 |
| Prostate | HPA | 0.4 |
| Spleen | HPA | 0.4 |
| Kidney | HubMAP | 0.229 |
| Large intestine | HubMAP | 0.7562 |
| Lung | HubMAP | 0.4945 |
| Prostate | HubMAP | 6.263 |
| Spleen | HubMAP | 0.4945 |

Finally, the images also differ in tissue section thickness. While all HPA images were sliced with a fixed thickness of 4 μm, the HuBMAP samples have tissue slice thicknesses ranging from 4 μm for the spleen and up to 10 μm for the kidney [1], adding another layer of complexity. The training dataset contained 352 samples along with additional metadata, including the dataset label (HPA or HuBMAP), organ, image height, image width, pixel size, tissue thickness, age (patient age), and sex (patient sex) [1]. During the testing stage, we had access to all meta information listed in the train dataset except for age and sex [1]. The test data comprised 550 images, of which 45% were for the public dataset and 65% for the private dataset [1]. The counterplot in Figure 3 also illustrates the class imbalance across organs, as presented in the training data.

## 4. Metric and Evaluation

For model evaluation Dice coefficient [31, 32] was used, which was simply averaged across all segmentation masks. For model evaluation metrics on three different datasets were used:
1. **Out Of Fold predictions**, using 5 Cross-Validation folds [33]. In order to preserve class imbalance and make metric more robust, stratification by organ was used.
2. Results from the **public Kaggle test only on the HubMAP part**. While the public Kaggle test set score was computed using both HPA and HuBMAP images, the final private dataset score was calculated using only HuBMAP data. We thus decided to focus solely on the HuBMAP score by not predicting masks for HPA images and adjusting the Kaggle public dataset score by the proportion of HuBMAP images (roughly 72%) [1].
3. Results from the **private Kaggle test set.**

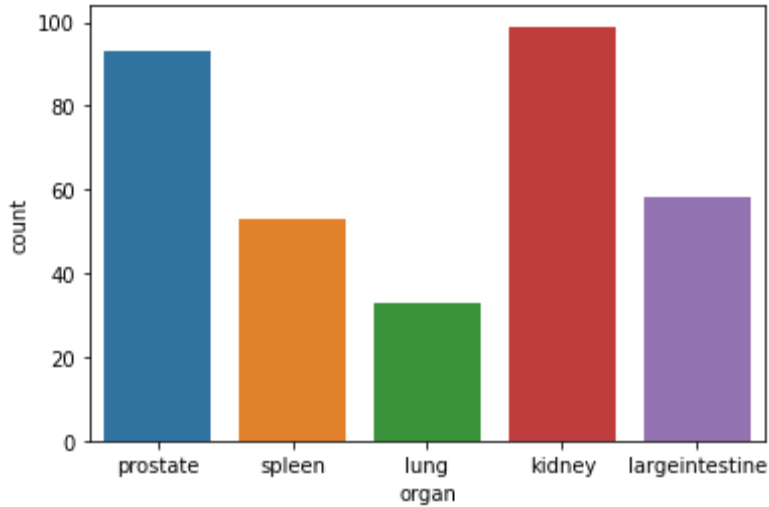

**Figure 3**: Class distribution of different organs in train dataset (HPA)

## 5. Methods
### 5.1. Model Architecture

We have used Unet [34] and Unet++ [35] architectures with pre-trained EfficientNet B7 [36] and Mix Vision Transformer [37] encoders. In our experiments, Unet++ showed comparable or better results compared to the pure Unet decoder, and Mix Vision Transformer outperformed EfficientNet B7 encoders on both Cross-validation and on Private Kaggle Dataset. For our final solution, we used a simple average of predictions from 15 models using EfficientNet B7 and Mix Vision Transformer encoder along with Unet and Unet++ style decoder which outperformed either of the single models (Table 5).

### 5.2. Data Preparation

In this challenge, competitors were asked to build a solution that can segment FTUs in a way that is invariant to the staining protocol (HPA or HubMAP). To achieve this goal, organizers provided competitors with image data for microscope slides stained using HPA protocol and evaluated solutions on the mixed HPA+HubMAP dataset for Public Leaderboard and on the HubMAP dataset only for the Private Leaderboard. Therefore, the biggest challenge for this competition was domain adaptation from the HPA dataset to HubMAP. In order to solve it we had to adapt our training data in 3 ways:
- Pixel size
- Color space difference
- Tissue thickness difference

**Adopting pixel size.**

One of the key points was adapting to wildly varying pixel sizes. The image scales ranged from 6.3um/pixel for the prostate to 0.2um/pixel for the large intestine. We tackled this issue by rescaling our train dataset to the target HuBMAP resolution. However, to increase the model's receptive field we applied additional downscalers for larger images and upscalers for smaller images (prostate). It is important to note that additional downscalers were also used at the inference stage to avoid changing the train/test pixel size. We used two datasets: one rescaled to HuBMAP scales and another with the original HPA scales. The latter one was not only important for HPA predictions (absent in the private LB) but also to provide some additional scaling information to the model. Therefore, we scaled down images of each organ by N times in order to match HubMAP pixel size and then by M times to upscale too small images of organs. Values of N and M can be found in Table 2.

14

**Table 2**
Scale sizes for each organ

| Organ | N | M |
|---|---|---|
| Kidney | 1.25 | 2 |
| Large intestine | 0.5725 | 2 |
| Lung | 1.8905 | 1 |
| Prostate | 15.65 | 0.3 |
| Spleen | 1.23625 | 2 |

**Adopting color space.**

The color spaces between HPA and HubMAP datasets were also different due to different stain methods - DAB [28] for HPA, PAS [29], and H&E [30] for HubMAP (see Figure 4). As the competition required segmentation of FTUs on slides stained using different staining protocols, we decided to make the neural network invariant to color variations by applying heavy color augmentations such as histogram matching [38] to match the color distribution of the training images to that of HuBMAP dataset (Figure 5).

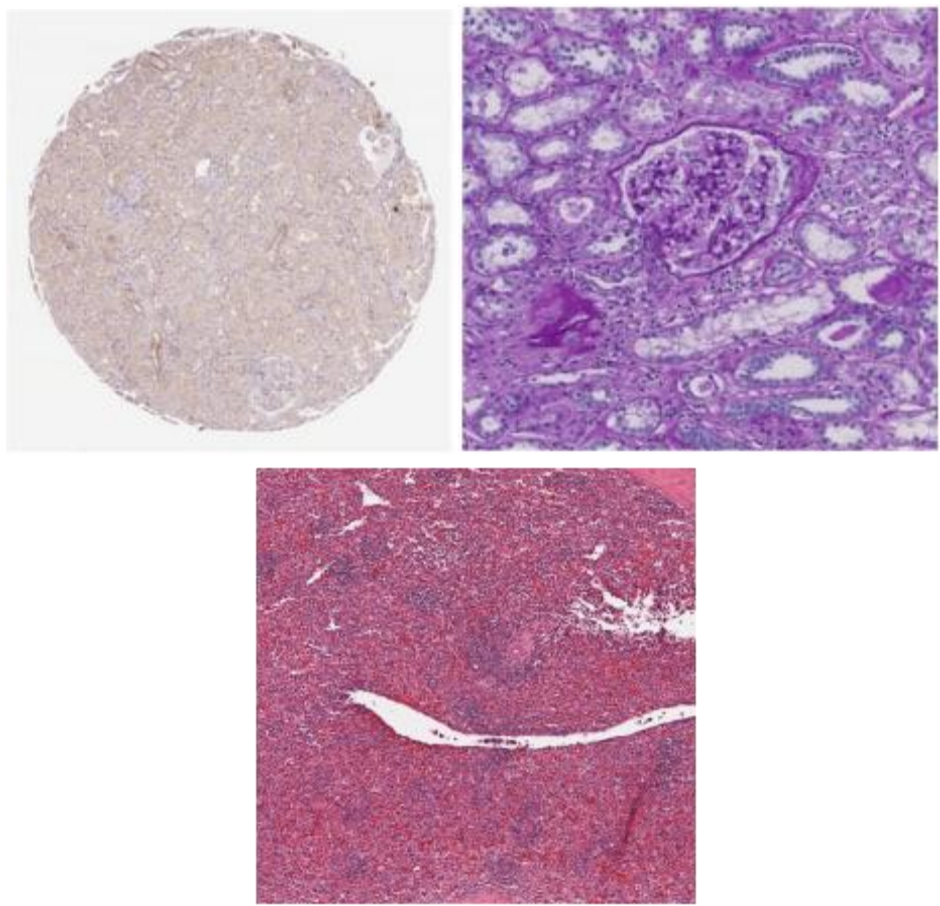

**Figure 4**: Examples of different stain methods. Top left - DAB, top right - PAS, bottom - H&E

We also applied hue-value-saturation, contrast, and gamma augmentations. To provide additional robustness to scale and geometrical differences in FTU shapes, we also applied a range of geometric augmentations, which included random flips, rotations, scales, shifts, elastic transforms, and more. Some competition participants chose to apply stain normalization [39] to cycle color between different staining protocols. However, in our experiments, we didn't see any improvement from stain normalization, probably because regular stain normalization techniques are specialized for one particular type of stain and don't work well when applied to images stained with different protocols.

We have gathered additional data from the GTEX portal [40] and a few images from HubMAP to which we applied histogram matching [38] of all train data to GTEX and HubMAP images. The results of histogram matching may be observed in Figure 5. Besides we have used heavy augmentations - Geometric, Color, Distortions, and Scales. The main idea behind the color augmentation was to suggest to the model that the color is not important and that it had to look for other features.

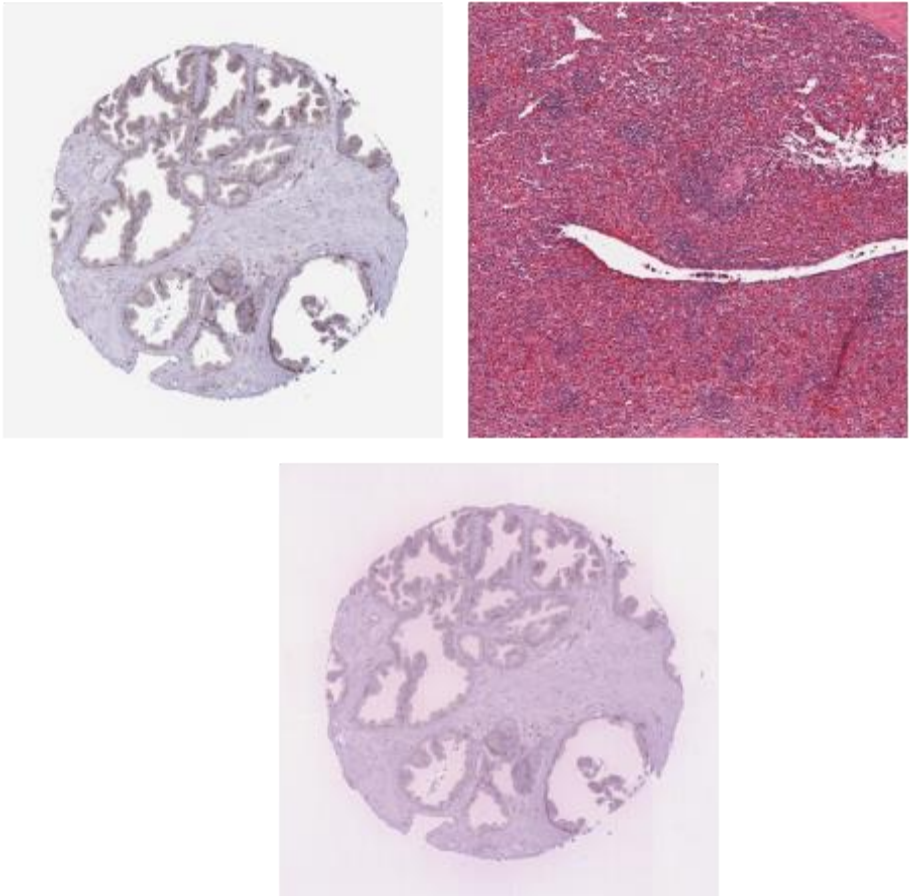

**Figure 5**: Left top - original spleen HPA image. Left right - original spleen HubMAP image. Bottom- matched HPA to HubMAP image

**External data.**

We have not tried to solve the problem of tissue thickness explicitly but we have decided to download additional data from different data sources and apply pseudo-labeling. We used data from GTEX [40] and HPA [19-26] portals to complement the initial training data. The GTEX data was especially important here because it was stained similarly to HuBMAP [27] slides with H&E [30]. From GTEX we downloaded prostate, large intestine, kidneys, and spleen data for patients with no apparent pathologies. We ignored lungs from GTEX as we couldn't figure out how to segment them and neither manually nor using pseudo labeling. We were progressively adding GTEX images to our pipeline ending up with around 140 at the end of the competition, though it is worth mentioning that each image was quite large measuring tens of thousands of pixels in width and height. From the HPA site, we used a plethora of DAB [28] stained slides very similar to those provided by organizers. Overall, we have added between 57-61K of additional HPA images for each organ.

We pseudo-labeled both HPA and HuBMAP images with the best ensemble (according to the Cross Validation Score) available at the time of labeling. We did not select the most confident pseudo labels but rather sampled the HPA and GTEX datasets at random at training time. The selection process was inspired by the pseudo-labeling technique proposed in a semi-supervised paper [41]. We have repeated the pseudo-labeling procedure twice. Examples of pseudo-labeled images can be observed in Figure 6.

**Cutmix.**

CutMix [42] augmentation was among the top contributors to our score. We applied it with a probability of 0.5 and used uniform distribution to sample which part of the original image to replace

16

with a patch from a different image. The key trick though was to apply CutMix augmentation within a single class. Examples of CutMixed image in Figure 7.

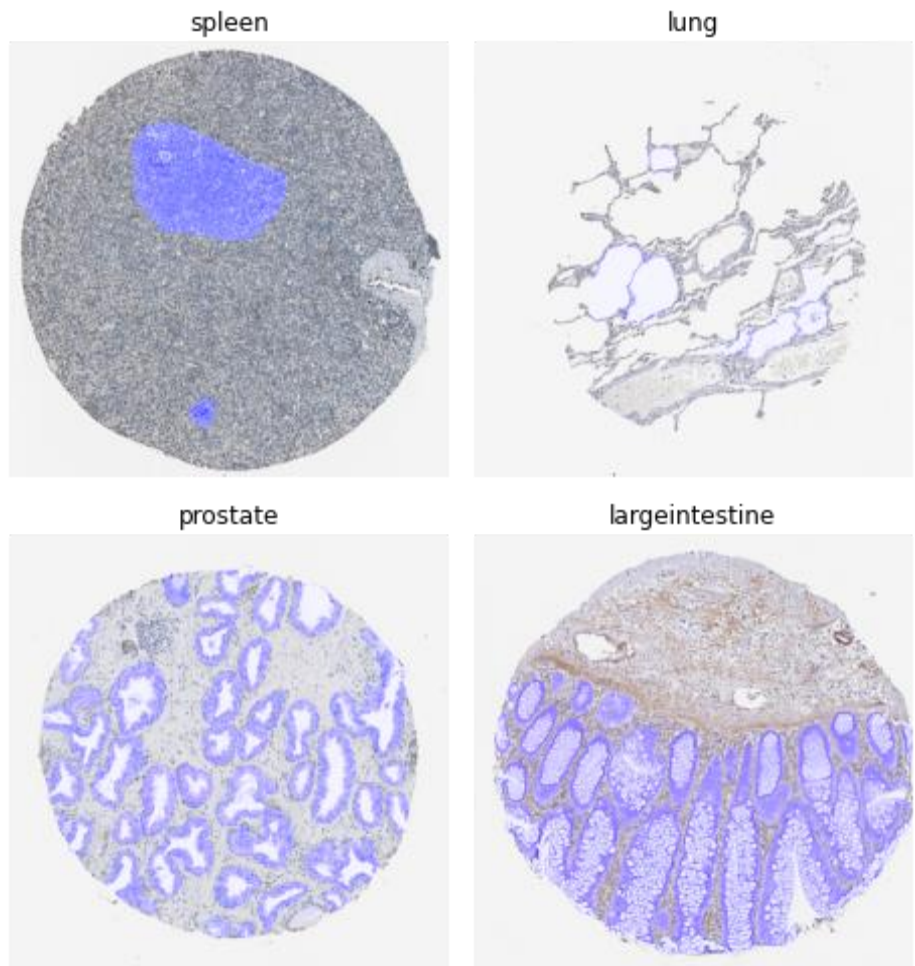

**Figure 6**: Examples of pseudo-labeled HPA images

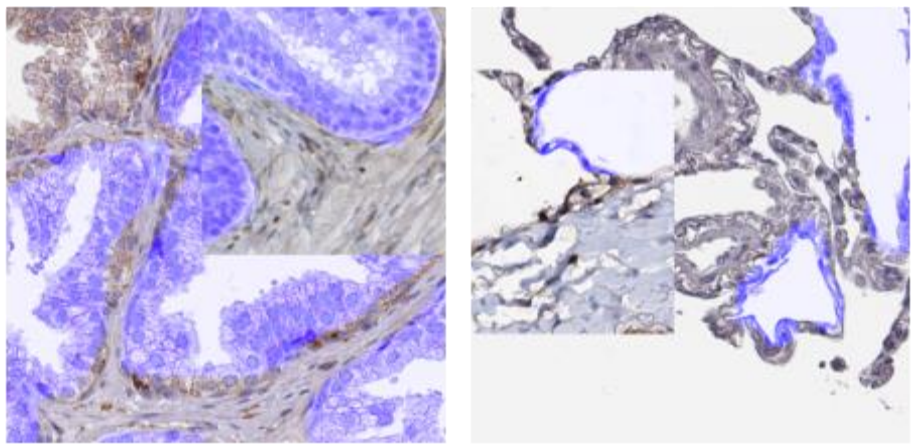

**Figure 7**: Examples of CutMixed HPA images (left - prostate; right - lung)

**Filtering Lung samples.**

FTUs on lungs were by far the most problematic part of the dataset with our baseline model scoring a mere 0.05 Dice on Cross Validation vs. 0.69 for the next hardest organ to segment - the spleen. Baseline model (Unet with EfficientNet B5 [34, 36]) Dice on different organs can be observed in Table 3.

**Table 3**
Cross Validation Dice for different organs

| Organ | Dice |
|---|---|
| Kidney | 0.85301 |
| Large intestine | 0.87770 |
| Lung | **0.04659** |
| Prostate | 0.81167 |
| Spleen | 0.69313 |

There were two major problems with lung FTUs (alveoli): first is inconsistent segmentation of the FTUs between images (Figure 8), and the second is the shortage of well-segmented samples. Alveoli on lung images were present in a collapsed and inflated form as well as horizontally and vertically sectioned. The horizontally sectioned inflated alveoli were the most abundant group, while collapsed and vertically sectioned images contained only 15 samples on our estimate. When used as a part of the train set, they generated too much noise and we decided to remove these samples from our training pipeline.

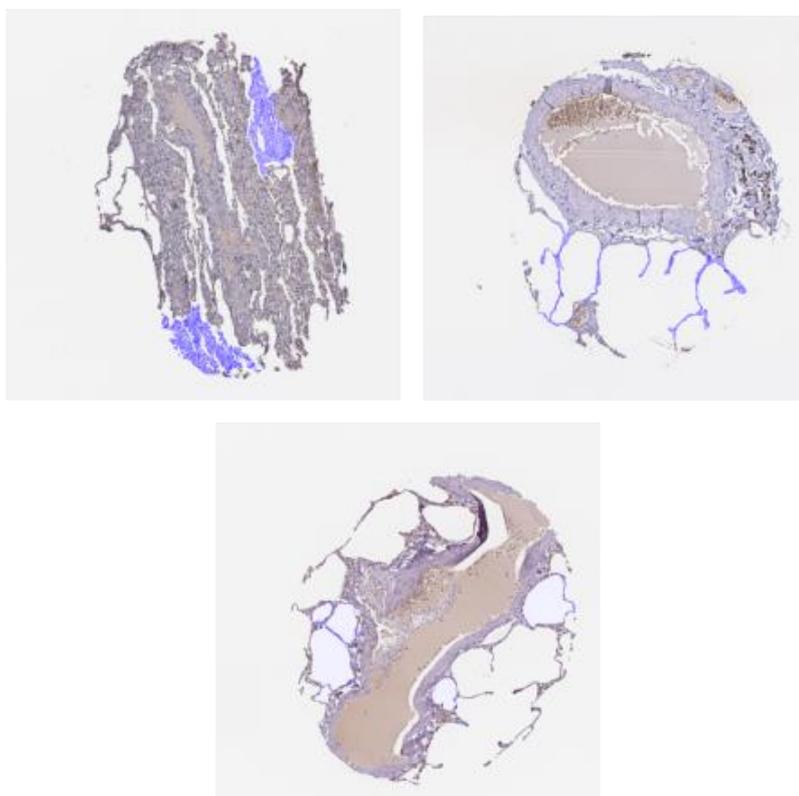

**Figure 8**: Example of lung FTU, highlighted with blue color

## 5.3. Training Process

We have used 512x512 training crops to train CNN models and 1024x1024 crops for Mix Vision Transformer [37] models. Non-empty masks sampled with 0.5 probability. For parameter optimization, we have used Adam optimizer [43] with an initial learning rate of 0.001. We reduced it in the training process with the help of the ReduceLROnPlateau [44] algorithm with patience 3 by 0.5 factor monitoring validation dice loss. Initially, the constructed pipeline was a multiclass model with 5 channels - one for each organ. However, as only one class of organ FTUs was present at any given image we reformulated the task as a binary semantic segmentation with a single channel containing all the masks no matter what organ was present on an image. Such an approach allowed for improved generalization and better scores across all organs. To improve model robustness we have used a mixture of four losses: binary Cross-Entropy, Dice Loss, Focal Loss [45], and Jacquard Loss [46]. We have also

18

trained an EfficientNet model with PointRend head [47] and scaled loss with a factor of 2. While we didn't notice a meaningful performance boost from the PointRend alone we think that its main contribution was in adding diversity to our model ensemble as well as some regularization. We have used PyTorch [48] built-in mixed precision training in order to reduce GPU memory consumption which allowed us to use a batch size of 32 samples on A100 GPUs.

### 5.4. Inference Process

For each fold of each of our final models we have averaged model parameters of 3 best checkpoints by validation dice [49]. For ensembling, we have simply averaged probability masks from each model. We have also used Test Time Augmentations [50] with original images and three flips. We have removed small regions after thresholding to reduce noisy masks. To do so we have used the next heuristic:

$$RegionArea/ImageArea < OrganTresh \quad (1)$$

OrganTresh for different organs was found empirically by testing its effects on Cross Validation Dice and can be found in Table 4.

**Table 4**
Relative Min Region area

| Organ | Dice |
|---|---|
| Kidney | 0.001 |
| Prostate | 0.0005 |
| Large intestine | 0.0001 |
| Spleen | 0.001 |
| Lung | 0.000001 |

We have used a 1024x1024 sliding window approach with 0.75 overlap for Mixed Vision Transformers [22] due to GPU memory constraints and predicted on full-scale images for CNN models.

## 6. Results
### 6.1. Final results

The results of training 5 models using 5 folds on out-of-fold data, public and private datasets are outlined in Table 5. Experiment ensembles include 5 models from each experiment and metrics from them are outlined in Table 6. Results of our approach compared to other top 5 best solutions can be found in Table 7.

**Table 5**
5 Folds Dice results

| Model | Out Of Fold | Public Leaderboard | Private Leaderboard |
|---|---|---|---|
| Unet ++ w/EfficientNet B7 | 0.84338 | **0.61189** | 0.82878 |
| Unet w/EfficientNet B7 | 0.83915 | 0.61160 | 0.82698 |
| Unet w/Mix Vision Transformer B5 | **0.85405** | 0.60826 | **0.83332** |
| Unet w/Mix Vision Transformer B3 | 0.85356 | 0.60273 | 0.82657 |

Analyzing results from both Table 1 and Table 2 we can make the following conclusions:
● Completely CNN approaches worked better on the Public HubMAP dataset, which can be caused by slight overfit to this test data.

- Mixed Vision models [37] outperformed CNN models both on Cross-Validation and Private test data, which can advise that these models perform better in terms of segmentation quality and domain adaptation.
- Mean ensemble of CNNs and Mixed Vision models [37] slightly improved results comparing to solo CNN or Mixed Vision Transformer [37] approach.

**Table 6**
Ensemble Dice results

| Model | Out Of Fold Dice | Public Leaderboard | Private Leaderboard |
|---|---|---|---|
| Unet ++ w/EfficientNet B7; Unet w/Mix Vision Transformer B5; Unet w/EfficientNet B7 X 2 | 0.84937 | **0.61453** | 0.83266 |
| Unet ++ w/ EfficientNet B7; Unet w/Mix Vision Transformer B5; Unet w/Mix Vision Transformer B3 | **0.85428** | 0.60931 | **0.83419** |

**Table 7**
Kaggle Top 5 Private results

| Best Results | Private Dice |
|---|---|
| First Place | 0.83562 |
| Second Place | 0.83393 |
| Unet ++. Encoder EfficientNet B7; Unet. Encoder Mix Vision Transformer B5; Unet. Encoder EfficientNet B7 X 2 (Our) | **0.83266** |
| Fourth Place | 0.82717 |
| Fifth Place | 0.82595 |

### 6.2. Ablation Study

In this section we will outline model performance improvements in terms of Dice score [31, 32] when we have introduced changes, described in previous sections - Table 8.

**Table 8**
Ablation study

| Head 1 | Out Of Fold Dice | Public HubMap Dice | Private Dice |
|---|---|---|---|
| Baseline | 0.71172 | 0.12115 | 0.15632 |
| + Pixel size adaptation | 0.71588 | 0.22653 | 0.30195 |
| + hist matching | 0.70283 | 0.38732 | 0.48486 |
| + 1 output channel + CutMix | 0.75157 | 0.49483 | 0.64518 |
| + Heavy augmentations | 0.76950 | 0.51506 | 0.71187 |
| + additional scalers + GTEX pseudo | 0.82142 | 0.58037 | 0.78352 |
| + HPA pseudo | 0.83633 | 0.59800 | 0.81402 |
| Best 5 Folds solo model | 0.85405 | 0.60826 | 0.83332 |
| Ensemble | 0.85428 | 0.60931 | 0.83419 |

From this table above we can clearly see that:

- Introduced changes improved Out of Fold Dice and Private Dice, which means that overall model performance increased both on HPA and HubMAP datasets.
- Introduced changes decreased, mostly eliminated the gap between Out of Fold Dice, and Private Dice, which means that they have completed the domain adaptation task between HPA and HubMAP datasets.

Also, each organ dice improved, especially the lung dice was improved more than 10 times - Table 9.

**Table 9**

Organ Dice Score comparison

| Model | Kidney | Large intestine | Lung | Prostate | Spleen |
|---|---|---|---|---|---|
| Baseline | 0.85301 | 0.87770 | 0.04659 | 0.81167 | 0.69313 |
| Best ensemble | 0.95677 | 0.91755 | 0.48833 | 0.83874 | 0.84872 |

## 7. Conclusion

This paper introduced the FTU segmentation training pipeline, which showed near state-of-the-art performance both on HPA [19-26] and HubMAP [27] datasets, minimizing the domain gap between them. Proposed methods allowed the adoption of models from the HPA domain to HubMAP, reducing the difference in the Dice score between test sets on HPA and HubMAP domains. Also, we have considerably increased our score on the HPA test set. We believe that the proposed methods can be used both for increasing the performance of semantic segmentation models on one domain and for adopting these models from one domain to another.

## 8. Acknowledgements

First, we would like to thank the Armed Forces of Ukraine, Security Service of Ukraine, Defence Intelligence of Ukraine, State Emergency Service of Ukraine for providing safety and security to participate in this great competition, complete this work, and help science, technology not stop and move forward. Also, we want to thank the Kaggle team, Google team, Genentech, and Indian University for hosting HuBMAP + HPA - Hacking the Human Body competition, which gave us all the needed data and materials to build models, test hypotheses, and write this paper.